\documentclass[epj]{svjour}
\usepackage{graphicx,amsmath,ifpdf,multirow}
\begin{document}

\title{Spectra, elliptic flow and azimuthally sensitive HBT radii 
	from Buda-Lund model for  $\sqrt{s_{NN}} = 200$ GeV Au+Au collisions 
	}
\titlerunning{Azimuthally sensitive HBT from Buda-Lund}
\author{
   Andr\'as Ster\inst{1,2}
   \and M\'at\'e~Csan\'ad\inst{3}
   \and Tam\'as Cs\"org\H{o}\inst{1,4}
   \and Bengt L\"orstad\inst{2}
   \and Boris Tom\'a\v{s}ik\inst{5,6}}
\institute{
MTA KFKI RMKI, H-1525 Budapest 114, PO Box 49, Hungary \and 
Department of High Energy Physics, University of Lund, S-22362 Lund, Sweden \and
E\"otv\"os University, H-1117 Budapest XI, P\'azm\'any P\'eter s. 1/A, Hungary \and
Department of Physics, Harvard University, 17 Oxford St, Cambridge, MA 02138, USA\and
Univerzita Mateja Bela, Tajovsk\'eho 40, SK-97401 Bansk\'a Bystrica, Slovakia \and
FNSPE, Czech Technical University in Prague, B\v{r}ehov\'a 7, CZ-11519 Prague, Czech Republic}
\authorrunning{Andr\'as Ster {\it et al.} }
\date{\today}

\abstract{We present calculations of elliptic flow and azimuthal
dependence of correlation radii in the ellipsoidally symmetric
generalization of the Buda-Lund hydrodynamic model of hadron production in high-energy 
nuclear collisions. 
We compare them to data from RHIC by simultaneous fits to azimuthally 
integrated invariant spectra of pions, kaons and protons-antiprotons
measured by PHENIX in Au+Au reactions at center of mass energy of 200 AGeV.
STAR data were used for azimuthally sensitive two-particle correlation 
function radii and for the transverse momentum dependence of the elliptic 
flow parameter $v_2$. 
We have found that the transverse flow is faster in the reaction plane then out of plane, which
results in a reaction zone that gets slightly more elongated in-plane than out of plane.
The model parameters extracted from the fits are shown and discussed. }


\maketitle

\section{Introduction}
\label{intro}

Important information about the properties of extremely hot strongly 
interacting matter comes from the observation of azimuthal anisotropies
in non-central ultra-relati\-vistic nuclear collisions. The second order Fourier  
component of azimuthal hadron distributions is connected with the azimuthal 
dependence of transverse 
collective expansion velocity of the bulk matter \cite{Heiselberg:1998es,Sorge:1998mk}.
That is in turn determined by the differences of the  initial pressure gradients
in the two perpendicular transverse directions, 
as well as by the initial geometry, 
the initial velocity and temperature distributions of the
fireball, and the equation of state~\cite{Heinz:2001xi,Csorgo:2001xm}. The anisotropic shape of the fireball 
measured with the help of correlation femtoscopy \cite{Wiedemann:1997cr} at the 
instant of final decoupling of hadrons bears information about the total lifespan 
of the hot matter: with time the originally out-of-reaction-plane shape becomes 
more and more round and may even become in-plane extended \cite{Heinz:2002sq}. Unfortunately, 
in determining the elliptic flow and azimuthally sensitive correlation radii 
individually two effects---spatial and flow anisotropy---are entangled. For example, 
the same elliptic flow can be generated with varying flow anisotropy strength
if the spatial anisotropy is adjusted appropriately~\cite{Tomasik:2004bn}.

In general, the precise way of the interplay 
between the two anisotropies is model dependent. It has been studied and shown 
to be different within the Buda-Lund model \cite{Csanad:2008af} as well 
as the Blast Wave model \cite{Tomasik:2004bn}. 

In this paper we analyze for the first time azimuthally sensitive Hanbury Brown -- Twiss (HBT) 
radii, using data from non-central heavy ion collisions within framework of the Buda-Lund model. 
Note that the model successfully describes data from central Au+Au collisions at RHIC,
as measured by BRAHMS, PHE\-NIX, PHOBOS, and STAR collaborations, 
including identified particle spectra and transverse mass dependent 
HBT radii as well as the pseudorapidity distributions of charged particles.
The model was shown before to describe the transverse mass and pseudorapidity
dependence of elliptic flow of identified particles at various energies and
centralities in ref.~\cite{Csanad:2005gv}. 
The Buda-Lund model formalism for non-central collisions, including
elliptic flow and azimuthal angle dependence of HBT radii has been 
proposed first in ~\cite{Csanad:2008af}. The model is defined  with the help of 
its emission function. In order to take into account the effects of resonance decays
it uses the core-halo model \cite{Csorgo:1994in}. In the present study, we improve
on earlier versions of the Buda-Lund model, by scrutinizing the various components
using azimuthally sensitive HBT data. 
Eventually we utilize a model that includes as a special case of T.S. Bir\'o 's  
axially symmetric and accelerationless exact solution of relativistic hydrodynamics~\cite{Biro:1999eh},
in contrast to the original, earlier variant, ref. \cite{Csanad:2008af}, which was based on
an ellipsoidally symmetric, but also non-accelerating exact solution of
relativistic hydrodynamics, given by ref. \cite{Csorgo:2003rt}.  
Similarly to ref. ~\cite{Csorgo:1999sj},
we present an improved calculation, using the binary source formalism,
to obtain the observables by using two saddle-points instead of only one. This results in
an oscillating pre-factor in front of the Gaussian in the two-pion correlation function
that we take into account for the formulae of the HBT radii.

Azimuthally sensitive HBT radii 
were also considered recently in cascade models, e.g. in the fast Monte-Carlo
model of ref.~\cite{Amelin:2007ic}, or, in the Hadronic Resonance Cascade~\cite{Humanic:2005ye}.

Data analysis of correlation HBT radii performed earlier with the Blast Wave model 
indicates that the fireball at the freeze-out is elongated slightly out of the reaction 
plane \cite{Adams:2004yc}, i.e. spatial deformation is similar as in the initial state given by the 
overlap function. This is also supported by the theoretical results from 
hydrodynamic simulations \cite{Heinz:2002sq,Frodermann:2007ab} and URQMD \cite{Lisa:2009wp}. 
It sets limitations on the total lifespan. From all previous analyses it seems, 
however, that the final state 
anisotropy has an interesting non-monotonous dependence on collision energy with a minimum 
at the SPS energies \cite{Lisa:2009wp}. In our analysis of the \emph{same} data with a 
\emph{different} model we observe for the first time at RHIC an in-plane elongation 
of the fireball at freeze-out.

The paper is structured as follows:
In Section~\ref{model} the basic features of the ellipsoidally symmetric 
Buda-Lund model are summarized. In Section~\ref{obs} we derive the analytic 
formulae for the observables such as elliptic flow and the azimuthally
 asymmetric correlation radii. In Section~\ref{fits} we show the results
of the simultaneous model fits to experimental data from non-central collisions
and we compare them to the ones obtained from fits to central data. 
In Section~\ref{conc} our conclusions are presented.

\section{Buda-Lund model: basic features}
\label{model}
We restrict ourselves here to a short description of the model, 
for details see refs.~\cite{Csanad:2008af,Csanad:2003qa}.

In the Buda-Lund model, the emission function is given by that of
a hydrodynamically expanding fireball (core), surrounded by a halo of
long lived resonances. The core emission function looks like:
\begin{align}\label{e:blsource}
S_c(x,p) d^4 x = \frac{g}{(2 \pi)^3} \frac{ p^\mu d^4\Sigma_\mu(x)}{B(x,p) +s_q},
\end{align}
where $g$ is the degeneracy factor ($g = 1$ for identified pseudoscalar mesons, 
$g = 2$ for identified spin=1/2 baryons), and $p^\mu d^4 \Sigma_\mu(x)$ is a 
generalized Cooper-Frye term, describing the flux of particles
through a distribution of layers of freeze-out hypersurfaces, $B(x,p)$ is the 
(inverse) Boltzmann phase-space distribution, and the term $s_q$ is determined 
by quantum statistics, $s_q = 0$, $-1$, and $+1$ for Boltzmann, Bose-Einstein 
and Fermi-Dirac distributions, respectively. Note that $x^\mu = (t,r_x,r_y,r_z)$
and $p^\mu = (E,p_x,p_y,p_z)$ are the four-vectors of the space-time point $x$
and the momentum $p$.  

For a relativistic, hydrodynamically expanding system, the (inverse) Boltzmann 
phase-space distribution is
\begin{align}
B(x,p)=\exp\left( \frac{ p \cdot u(x)}{T(x)} -\frac{\mu(x)}{T(x)}\right).
\end{align}
The forms of the flow four-velocity  ($u_\nu(x)$), chemical potential  ($\mu(x)$), and
temperature  ($T(x))$ distributions are introduced below. Note that it can be mapped onto 
exact solutions of hydrodynamics, both in the relativistic and in the non-relativistic
cases, as detailed in ref.  ~\cite{Csanad:2003qa}. For example, let us mention, that
in the non-relativistic limit, the Buda-Lund hydro model
corresponds to the exact, parametric, ellipsoidally symmetric
solutions of non-relativistic hydrodynamics in ref.~\cite{Csorgo:2001xm} 
which solution at late times converges to an accelerationless exact solution of relativistic hydrodynamics as detailed in ref.~\cite{Csorgo:2003ry}.
According to our best knowledge, no similar connection has been explored yet in case of the
azimuthally sensitive version of the Blast Wave model of ref.~\cite{Retiere:2003kf},
and known exact parametric solutions of (relativistic) hydrodynamics.

The generalized Cooper-Frye pre-factor, as described in refs.~\cite{Csanad:2008af,Csanad:2003qa} was
\begin{align}
\label{s:pmu}
p^\mu d^4 \Sigma_\mu(x) =p^\mu u_\mu(x) H(\tau) d^4 x.
\end{align}
The time dependence of the emission, described by $H(\tau)$ was
approximated with a Gaussian distribution around the freeze-out proper-time $\tau_0$,
\begin{equation}
H(\tau) = \frac{1}{(2 \pi \Delta\tau^2)^{1/2}}
\exp\left(-\frac{(\tau - \tau_0)^2 }{ 2 \Delta \tau^2}\right),
\end{equation}
with $\Delta \tau$ being the duration of the particle production in 
longitudinal proper-time ($\tau=\sqrt{t^2-r_z^2}$).
Of course, this function $H(\tau)$ can be easily generalized to have
more complicated forms, but the data discussed in the present
paper do not require us to go beyond the Gaussian approximation.
However, we found that the analysis of the azimuthally sensitive HBT radii was
actually sensitive to the structure of the Cooper-Frye pre-factor.
Eq.~(\ref{s:pmu}) corresponds to freeze-out hypersurface layers that are
pseudo-orthogonal to the four-velocity. For flow profiles with significant
longitudinal and radial flows, as specified below, these hypersurfaces have
positive correlations between the transverse radial coordinates $r_t$ and
time $t$. When defining the axially symmetric Buda-Lund model in ref. ~\cite{Csorgo:1999sj}, 
such positive $(r_t,t)$ correlations were neglected
and the freeze-out hypersurface was assumed to be a constant in the transverse direction.
Recently, new exact analytic solutions of relativistic hydrodynamics also lead to
freeze-out hypersurfaces with the property of nearly negligible $(r_t,t)$ correlations, see ref.~\cite{Gubser:2010ze}.
Based on favourable comparisons with data and analogies to the axially symmetric 
Buda-Lund model, we
decided to keep this kind of freeze-out hypersurfaces for the purpose of the present paper.
Note also that following ref.~\cite{Csorgo:1999sj} of the axially symmetric case, we
include a factor $\tau$ to $H_*(\tau)$ and approximate it by a Gaussian $H(\tau)$. 
Thus our modified Cooper-Frye term reads as
\begin{align}
p^\mu d^4 \Sigma_\mu(x) = m_t \cosh(\eta -y) H(\tau)   d\tau \, \tau_0 \, d\eta \, dr_x dr_y,
\end{align}
where $m_t=\sqrt{m^2+p_t^2}$ is the transverse mass, $p_t$ is the transverse momentum, $y$ is the rapidity and $\eta$ is the longitudinal space-time rapidity $\eta=0.5\log[(t+r_z)/(t-r_z)]$.
The four-velocity field, $u^\mu(x)$ assumed to be a directional Hubble flow, where
two different radial Hubble constants ($H_x$ and  $H_y$) characterize the different strength of the flow in the
impact parameter plane ($x-z$) and out of this plane ($y-z$). This transverse flow is assumed to develop
on a flow profile that is Bjorken-type at the $r_t = 0$ axes of the collision :
\begin{align}
u^\mu(x)& = (\cosh[\eta] \cosh[\eta_t] , H_x r_x, H_y r_y, \sinh[\eta] \cosh[\eta_t])
\end{align}
with $H_x = {\dot R}_x / R_x$  and $H_y ={\dot R}_y / R_y$ are the
defining relationships for the flow in the impact parameter plane (called also as reaction plane) and in the remaining orthogonal transverse direction. $R_x$ and $R_y$ are the characteristic 
geometrical system sizes in the two transverse directions, whereas $\dot R_x$ and $\dot R_y$
are their time derivatives.
The average transverse fluid rapidity $\eta_t$ is also introduced
with the defining relation $\sinh^2[\eta_t] = r_x^2 H_x^2 +  r_y^2 H_y^2$,
which ensures that $u\cdot u = 1$. Note that at mid-rapidity,  with $\eta = 0$,
the velocity profile of T.S. Bir\'o 's  axially symmetric and accelerationless exact solution of relativistic hydrodynamics~\cite{Biro:1999eh}
corresponds to the ${\dot R}_x / R_x = {\dot R}_y / R_y$ case, while the
solution discussed in ref. ~\cite{Csorgo:2003rt} coincides with this velocity profile.

For the fugacity distribution $\exp\frac{\mu(x)}{T(x)}$ we assume
\begin{equation}
\frac{\mu(x)}{T(x)} = \frac{\mu_0}{T_0} - 
\frac{r_x^2}{2R_x^2} - \frac{r_y^2}{2R_y^2} -\frac{(\eta - y_0)^2}{2\Delta\eta^2},
\end{equation}
so that it leads to a Gaussian in coordinate space. 
$\Delta \eta$ denotes the space-time rapidity width and $y_0$ is the mid-rapidity.

For the temperature profile we use the following form:
\begin{equation}
\frac{1}{T(x)}= \frac{1}{T_0} 
\left( 1 + a_x^2 \frac{r_x^2}{2R_x^2} + a_y^2 \frac{r_y^2}{2R_y^2} \right)
\left( 1 + a_{\tau}^2 \frac{(\tau -\tau_0)^2}{2 \Delta\tau^2}\right),
\end{equation}
where $T_0$ is the temperature of the collision center 
at the mean freeze-out time $\tau_0$. The parameters $a_x$, $a_y$ and $a_{\tau}$ control
the transversal and the temporal changes of the local temperature profile.
Note that its dependence on the coordinate $\eta$ will not be studied here 
(i.e. $a_\eta=0$ is assumed). 

\section{Observables from the Buda-Lund model}
\label{obs}
The observables can be calculated analytically from the Buda-Lund hydro model, 
using a double saddle-point approximation in the integration. 
We quote results from ref. \cite{Csanad:2003qa}. 
Note that in the binary source formalism of
ref.~\cite{Csorgo:1999sj} (Section 8 and 9) the double saddle-points are generated from the saddle-point $\overline{x}$ of the Boltzmann term by the product with the Cooper-Frey pre-factor in which the two exponentials in the pre-factor generate two terms with separate saddle points. Hence, in the final formulae only $\overline{x}$ appears.  
The saddle point coordinates 
$\overline{x}^\mu=$ ($\overline{\tau}\cosh(\overline{\eta})$,$\overline{r}_x$,$\overline{r}_y$, 
          $\overline{\tau}\sinh(\overline{\eta}) $) 
and the longitudinally boost invariant average emission widths 
($\Delta \overline{\tau}$, $\Delta \overline{\eta}$, $\overline{R}_x$, $\overline{R}_y$) 
are given as:
\begin{subequations}
\begin{align}
\overline{\tau} = &\;\tau_0, \\
\overline{\eta} = &\frac{y_0 - y} {1 + \Delta \eta^2 m_t / T_0} + y, \\
\overline{r}_i  = &\frac{p_i \dot R_i R_i / T_0}{1 + (a_i^2 + {\dot R}_i^2) \overline{E}/T_0}\;
\mathrm{ for }\;i=x,y,\\
\nonumber \\
\Delta \overline{\tau}^2 = & \frac{\Delta \tau^2} {1 + a_{\tau}^2     \overline{E}/T_0}, \\
\Delta \overline{\eta}^2 = & \frac{\Delta \eta^2} {1 + \Delta \eta^2  \overline{E}/T_0}, \\
\overline{R}_i^2         = & \frac{R_i^2} {1 + (a_i^2 + {\dot R}_i^2) \overline{E}/T_0}) \;
\mathrm{ for }\;i=x,y, \label{e:rstar} \\
\end{align}
\end{subequations}	
where $\overline{E} =  m_t \cosh(\overline{\eta} - y)$.
The invariant momentum distribution is evaluated using an ellipsoidally symmetric
generalization of eqs. (127, 130-140) of ref.
~\cite{Csorgo:1999sj}, that were first derived for the case of axially symmetric collisions:
\begin{align}\label{e:n1p}
E \frac{d^3N}{dp^3} = {N_1}(p)= \frac{g}{(2 \pi)^3} \overline{E}  \; \overline{V}\; \overline{C} \; \frac{1}{B(\overline{x},p)+s_q},
\end{align}
where
\begin{align}
\overline{V} & = (2 \pi)^{3/2}\, \frac{\Delta \overline{\tau}}{\Delta \tau}\, \overline{R}_\parallel \overline{R}_x \overline{R}_y,\\
\overline{C} & = \frac{1}{\sqrt{\lambda_*}} \exp(\Delta \overline \eta^2 / 2).
\end{align}
In the latter two expressions we use the notation $\overline{R}_\parallel = \tau_0 \Delta \overline \eta$  and  $\lambda_*$ is the ($y$ and $p_t$ dependent) intercept parameter of the two-particle correlation function.

The axially symmetric limit of these formulae, given in ref.~\cite{Csorgo:1999sj}
corresponds to the replacements of 
$H_x=H_y \rightarrow H_t$, 
$a_x=a_y \rightarrow a_r$, 
$r_x \rightarrow r_t$, $r_y \rightarrow 0$ and
$\overline{R}_x \overline{R}_y \rightarrow \overline{R}_t^2$.

The azimuthal angle ($\phi$) dependence of the invariant momentum distribution, eq.~(\ref{e:n1p}), can be re-expressed using
a Fourier-expansion in $\phi$:
\begin{align}
\label{e:n1pt0}
N_1(p)=N_1(p_t, p_z)\left[1+2\sum_{\substack{n=1}}^{\infty} v_n \cos(n\phi)\right],
\end{align}
where $v_n$ are the flow coefficients, in particular $v_2$ is the elliptic flow.
Sine terms do not appear in the expression due to mirror symmetry with respect to 
the reaction plane.  
The azimuthally averaged transverse momentum distribution $N_1(p_t, p_z)$ from $N_1(p)$ of eq.~(\ref{e:n1p}) is
\begin{align}
N_1(p_t,p_z)=\frac{1}{2\pi}\int_{0}^{2\pi} N_1(p) d\phi
\end{align}
and the elliptic flow:
\begin{align}
v_2(p_t,p_z)=\frac{\int_{0}^{2\pi} d\phi N_1(p) \cos(2\phi)}{\int_{0}^{2\pi}d\phi N_1(p)}.
\end{align}
At mid-rapidity ($p_z = 0$), in reactions of equal-mass nuclei the terms of odd coefficients in eq.~(\ref{e:n1pt0}) disappear due to the symmetry $\phi \rightarrow \phi + \pi$.    
In this particular case,
this observable can easily be expressed analytically if we assume that $v_2 \gg v_n$ if $n \geq 4$, which we may conclude from data~\cite{Adare:2010vv} where $v_4  \sim v_2^2  \sim 0.01$ or smaller, 
hence it can be neglected. 
Then the formula in eq.~(\ref{e:n1pt0}) simplifies to
\begin{align}
N_1(p_x,p_y,p_z=0) = N_1(p_t) \left[ 1 + 2 v_2 \cos(2 \phi) \right],
\end{align}
where $p_t = \sqrt{p_x^2 + p_y^2}$ and $p_x=p_t \cos \phi, p_y=p_t \sin \phi$. If we evaluate this equation at two appropriate angles ($\phi = 0^o$ and $\phi=45^o$) then
the transverse momentum dependence of the coefficient of the elliptic flow can be re-expressed
as follows
\begin{align}
v_2(p_t)=\frac{1}{2} \left( \frac{N_1( p_t, 0, 0)}{N_1 \left( \frac{p_t}{\sqrt{2}}, \frac{p_t}{\sqrt{2}}, 0 \right) } - 1\right).
\end{align}
This leads to a simple analytic derivation of $v_2(p_t)$ at mid-rapidity. 
Numerical investigations at the physical values of the model parameters 
indicate that the formulae have about 1\% relative error, only. 
Note that the azimuthally averaged transverse momentum distribution in this case 
takes also a simple form as
\begin{align}
N_1(p_t)= N_1 \left( \frac{p_t}{\sqrt{2}}, \frac{p_t}{\sqrt{2}}, 0 \right).
\end{align}

 However, we can use another but less precise method, as well, to find analytic approximation for the elliptic flow. In the next procedure a scaling variable $w$ is introduced. We show that after some approximations, $v_2$ depends on any variable through this variable $w$ only, hence $v_2(w)$ is a universal function as already pointed out in refs.~\cite{Csorgo:2001xm,Csorgo:1994in,Csanad:2003qa,Csanad:2005gv}. Both methods were tested against data but the previous one proved to describe them with better confidence.

If we evaluate $B(\overline{x},p)$ at mid-rapidity
in the limit, where the saddle-point coordinates are all small, we get:
\begin{equation}
\ln B(\overline{x},p)= \frac{p_x^2}{2 m_t T_x} + \frac{p_y^2}{2m_t T_y} - \frac{p_t^2}{2m_t T_0} + \frac{m_t}{T_0} - \frac{\mu_0}{T_0},
\end{equation}
where the direction dependent slope parameters are
\begin{eqnarray}
T_x&=&T_0+m_t \, {\dot R}_x^2 \frac{T_0}{T_0 + m_t a_x^2},\\
T_y&=&T_0+m_t \, {\dot R}_y^2 \frac{T_0}{T_0 + m_t a_y^2}.
\end{eqnarray}
The result for the (azimuthally integrated) transverse momentum spectrum is:
\begin{eqnarray}
\label{e:n1pt}
  N_1(p_t)  & \approx & \frac{g}{(2 \pi)^3} \left[\overline{E}  \overline{V} \overline{C}\right]_{p_x=p_y=p_t / \sqrt{2} \,\,} 
\exp\left[-\frac{p_t^2}{2 m_t T_{\rm eff}}\right] \nonumber \\
\end{eqnarray}
where we have introduced $T_{\rm eff}$, the effective slope of the azimuthally
averaged single particle $p_t$ spectra as the harmonic mean of the
slope parameters in the in-plane and in the out-of-plane transverse directions,
\begin{align}
\frac{1}{T_{\rm eff}} = \frac{1}{2} \left ( \frac{1}{T_x} + \frac{1}{T_y} \right )\, ,\label{e:teff}
\end{align}

The result for the elliptic flow is the following simple scaling law:
\begin{align}
\label{vteq}
v_2 \cong \frac{I_1(w)}{I_0(w)},
\end{align}
where $I_n(z)$ stands for the modified Bessel function of the
second kind, $I_n(z) = (1/\pi) \int_0^\pi \exp(z \cos\theta)
\cos(n \theta) d\theta $. The scaling variable
\begin{align}
w = \frac{p_t^2}{4m_t} \left ( \frac{1}{T_y} - \frac{1}{T_x} \right )\, .\label{e:w}
\end{align}
This can also be written as
\begin{align}
w = E_K\, \frac{\epsilon}{T_{\rm eff}}\, ,
\end{align}
where $E_K$ is a relativistic generalization of the transverse
kinetic energy, defined as
\begin{align}
E_K = \frac{p_t^2}{2 m_t}\, ,
\end{align}
and  momentum space eccentricity parameter
\begin{align}
\epsilon  =  \frac{T_x - T_y}{T_x + T_y}\,.
\end{align}

In order to compare more easily the Buda-Lund model results for the elliptic flow with the
azimuthally sensitive extension of the Blast Wave model \cite{Retiere:2003kf,Tomasik:2004bn},
we introduce $\rho_0$ and $\rho_2$ so that
\begin{subequations}
\begin{align}
{\dot R}_x & = \rho_0 (1 + \rho_2)\\
{\dot R}_y & = \rho_0 (1 - \rho_2)
\end{align}
\end{subequations}
therefore
\begin{subequations}
\begin{align}
\label{e:rho0}
\rho_0 & = \frac{1}{2} \left ( {\dot R}_x + {\dot R}_y \right )\\
\label{e:rho2}
\rho_2 & = \frac{{\dot R}_x - {\dot R}_y}{{\dot R}_x + {\dot R}_y}\, .
\end{align}
\end{subequations}

The elliptic flow in the Buda-Lund model also depends on the
transverse temperature gradients $a_x$ and $a_y$. The difference between them
actually moderates the difference between $T_x$ and $T_y$ and
modifies the elliptic flow at given  $\rho_0$ and  $\rho_2$. 
On the other hand, at larger momenta and
small temperature gradient, the following approximative proportionality holds:
\begin{align}
v_2 \propto \frac{\rho_2}{(\rho_2^2-1)^2}.
\end{align}

In the following we detail the results on HBT radii. 
We assume that  the fireball is not tilted in the reaction plane,
i.e.\ there is no $x$-$z$ correlation in the emission points of the 
hadrons\footnote{This is justified at RHIC and higher energies, $\vartheta = 0$. A tilt
with $\vartheta > 0$ may 
appear at lower energies.}.
There is, however, an angle $\varphi$ between the main axis of the ellipsoidal cross-section 
of the fireball and the outward and sideward axes given by the momentum of the hadrons. 
The former are determined by the orientation of the reaction plane while the latter 
are defined so that the outward axis agrees with the direction of the average transverse 
momentum of the pair and the sideward axis is perpendicular to it. 
Following eq. (128) in ref.~\cite{Csorgo:1999sj},
the formula for the two-particle Bose-Einstein correlation function can be expressed as 
\begin{equation}
C_2({\bf k_1, k_2}) = \frac{N_2(\bf k_1, k_2)}{N_1({\bf k_1}) N_1(\bf k_2)} 
= 1 + \lambda_* \Omega(q_\parallel) \exp(-q_i^2 \overline{R}_i^2), \\
\end{equation}
where $q=k_1 - k_2=(q_{=},q_\parallel,q_x,q_y)$, 
$\overline{R}=(\Delta \tau, \overline{R}_\parallel, \overline{R}_x, \overline{R}_y)$
and in the exponential we use the Einstein summation rule over the same indices.
The longitudinally boost invariant $temporal$, $parallel$, $side$ward, $out$ward 
relative momentum components are defined as
\begin{subequations}
\begin{align}
q_{=}         \, & = \, q_0 \cosh(\overline{\eta}) - q_z \sinh(\overline{\eta}) ,\\
q_{\parallel} \, & = \, q_z \cosh(\overline{\eta}) - q_0 \sinh(\overline{\eta}) ,\\
q_{out}       \, & = \, (q_x K_y - q_y K_x) / \sqrt{K_x^2 + K_y^2},\\
q_{side}      \, & = \, (q_x K_x + q_y K_y) / \sqrt{K_x^2 + K_y^2},\\
Q^2              & =-q^\mu q_\mu \, 
                   = - q_{=}^2 + q_{\parallel}^2 + q_{x}^2 + q_{y}^2 \\
                 & = - q_{=}^2 + q_{\parallel}^2 + q_{side}^2 + q_{out}^2, 
\end{align}
\end{subequations}
where $K = 0.5(k_1+k_2)$. 
The pre-factor $\Omega$ induces oscillations within the Gaussian envelope as a function 
of $q_\parallel$. This factor is given as
\begin{equation}
\Omega(q_{\parallel}) = 
\cos^2(q_{\parallel} \overline{R}_\parallel \Delta \overline \eta) + 
\sin^2(q_{\parallel} \overline{R}_\parallel \Delta \overline \eta) \tanh^2(\overline{\eta}).
\end{equation}
However, it can be approximated by a Gaussian, too, hence we may merge it with the
longitudinal and temporal emission widths as:
\begin{subequations}
\begin{align}
\overline{R}_{\parallel,G}^{2}  & = 
R_\parallel^2 (1 + \Delta \overline{\eta}^2 / \cosh^2(\overline{\eta})),\\
\Delta \overline{\tau}_G^2      & = 
\sinh^2(\overline{\eta}) R_{\parallel,G}^{2}+\cosh^2(\overline{\eta}) \Delta \overline{\tau}^2.
\end{align}
\end{subequations}

A widely used parameterization of the correlation function is given in the 
longitudinally co-moving frame (LCMS \cite{Csorgo:1991kfki}, $\beta_l = 0$) 
of the particle pair and in the out-side-long system~\cite{Bertsch:1988db} 
of Bertsch-Pratt
(where $K_{\mu}=(K_0,K_{out},0,0)$):
\begin{equation}
C_2({\bf k_1, k_2}) \simeq 1 + \lambda_* \exp (- q_i q_j R_{ij}^2),
\end{equation}
where $i,j=(out,side,long)=(o,s,l)$.
Then the formulae for these HBT radii can be expressed by the following transformations
\begin{subequations}
\label{oslradii}
\begin{align}
R_{o}^2 & = \overline{R}_x^2 \cos ^2 \varphi + \overline{R}_y^2 \sin^2 \varphi 
          + \beta_{o}^2 \Delta \overline{\tau}_G^2 \\
        & = \frac{\overline{R}_x^2 + \overline{R}_y^2}{2} 
          + \beta_{o}^2 \Delta \overline{\tau}_G^2 
          - \frac{\overline{R}_y^2 - \overline{R}_x^2}{2} \cos(2\varphi)\, \nonumber \\
R_{s}^2 & =  \overline{R}_x^2 \sin ^2 \varphi + \overline{R}_y^2 \cos^2 \varphi\\\label{rside}
        & = \frac{\overline{R}_x^2 + \overline{R}_y^2}{2} 
          + \frac{\overline{R}_y^2 - \overline{R}_x^2}{2} \cos(2\varphi)\, \nonumber,\\
R_{os}^2& = \frac{\overline{R}_y^2 - \overline{R}_x^2 }{2} \sin (2\varphi)\, , \\
R_{l}^2 & = \cosh^2(\overline{\eta}) \overline{R}_{\parallel,G}^2 
          + \sinh^2(\overline{\eta}) \Delta \overline{\tau}^2\, ,\\
R_{ol}^2& = -\beta_{o} \sinh(\overline{\eta}) \cosh(\overline{\eta}) (\overline{R}_\parallel^2 
          + \Delta \overline{\tau}^2)\, , \\
R_{sl}^2 & = 0\, .
\end{align}
\end{subequations}


\section{Comparison to experimental data}
\label{fits}

We have determined the best parameter values by fitting
the analytic expressions for the observables, given in the previous section, 
to experimental data, 
with the help of the CERN Minuit fitting package. 
Data from 20-30\% centrality class of 200 AGeV Au+Au collisions provided by 
PHENIX~\cite{Adler:2003cb,Adler:2004rq} 
and STAR~\cite{Adams:2004bi,Adams:2003ra} were used in the analysis.
The fits were performed simultaneously to azimuthally integrated 
transverse mass spectra of positive and negative pions, kaons, and 
(anti)protons \cite{Adler:2003cb}, the transverse momentum
dependence of the elliptic flow parameter $v_2$ of pions~\cite{Adams:2004bi}
and to the HBT radii due to pion correlations
as functions of transverse mass and the azimuthal angle~\cite{Adams:2003ra}. 
The results are plotted in Figs. 1-5. 

The interpretation of the model parameters is summarized in Table~\ref{t:pars}. 
However, for a better understanding the results three alternative parameters are introduced
that can be expressed from the previous ones the following way:
\begin{subequations}
\begin{align}
R_{sx}^2 &= \frac{2R_x^2} {a_x^2}&\\
R_{sy}^2 &= \frac{2R_y^2} {a_y^2}&\\
T_{e}    &= \frac{T_0}    {1 + a_{\tau}^2}. &
\end{align}
\end{subequations}
The two radii parameters correspond to the thermal surface sizes where the temperature 
drops to $T_0/2$, and the parameter $T_e$ corresponds to the temperature of the center after
most of the particle emission is over (cooling due to evaporation an expansion). 
Sudden emission corresponds to  $T_e = T_0$, and the $\Delta\tau \rightarrow 0$ limit. 
Also note that we use $\mu_B$, baryochemical potential. This is calculated from the chemical potential of protons and antiprotons:
$\mu_B = 1/2 \  (\mu_{0,p}-\mu_{0,\overline{p}})$, as written in Table~\ref{t:pars}.

In Table~\ref{t:results}, we present the model parameters obtained from simultaneous 
fits to the data sets. For comparison, results are shown from our earlier 
analysis of 0-30\% centrality collisions~\cite{Csanad:2004mm}, too, 
that was performed with a previous version of the model corresponding to the 
axially symmetric limit of the current ellipsoidally generalized 
Buda-Lund hydrodynamic model.

\begin{figure}
	\centering
		\includegraphics[width=0.45\textwidth]{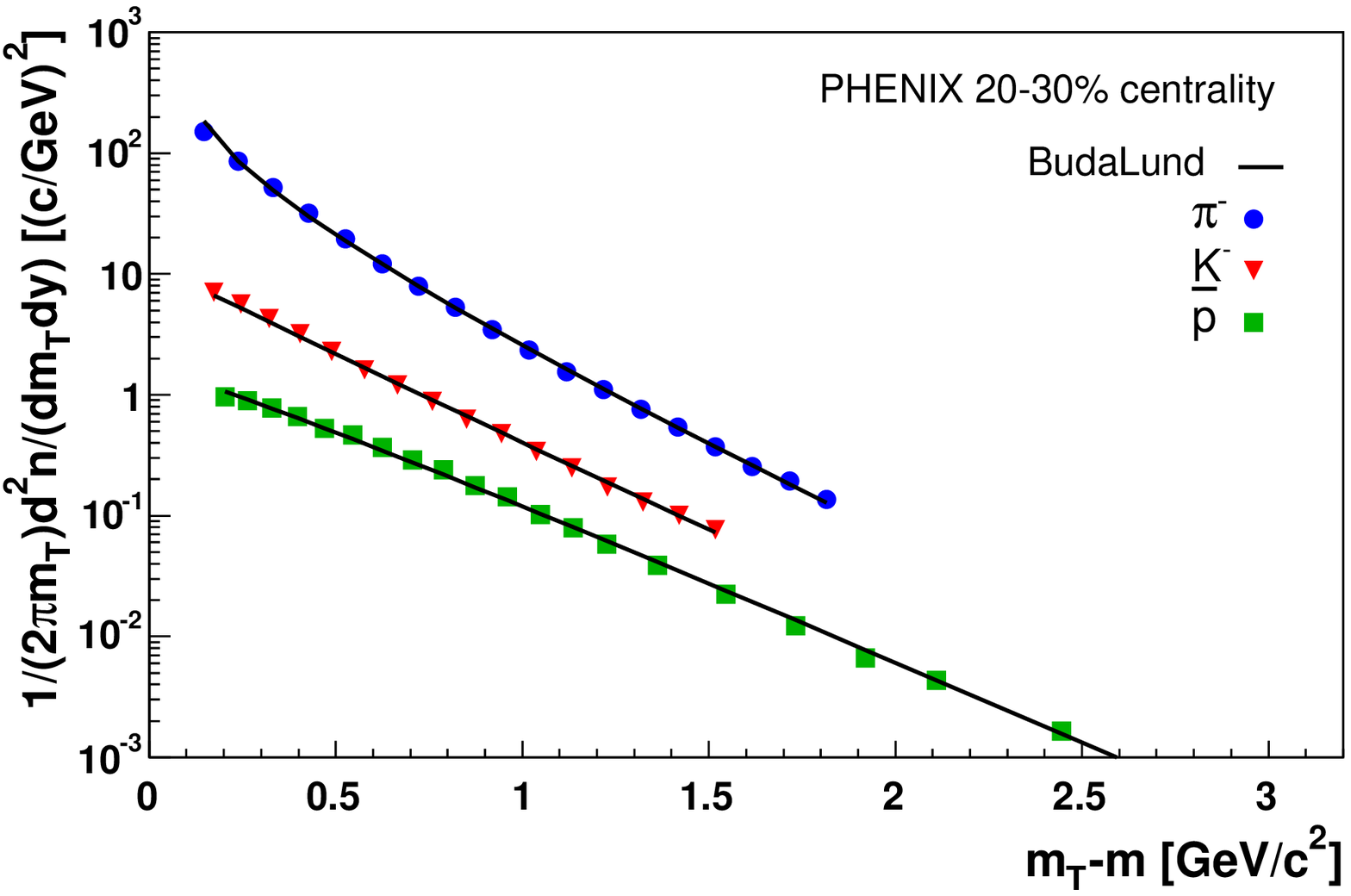}
\caption{Buda-Lund model fits to RHIC 200GeV Au+Au data
of azimuthally integrated transverse momentum spectra of 
negatively charged particles data~\cite{Adler:2003cb}.}
	\label{f:fig1}
\end{figure}
\begin{figure}
	\centering
		\includegraphics[width=0.45\textwidth]{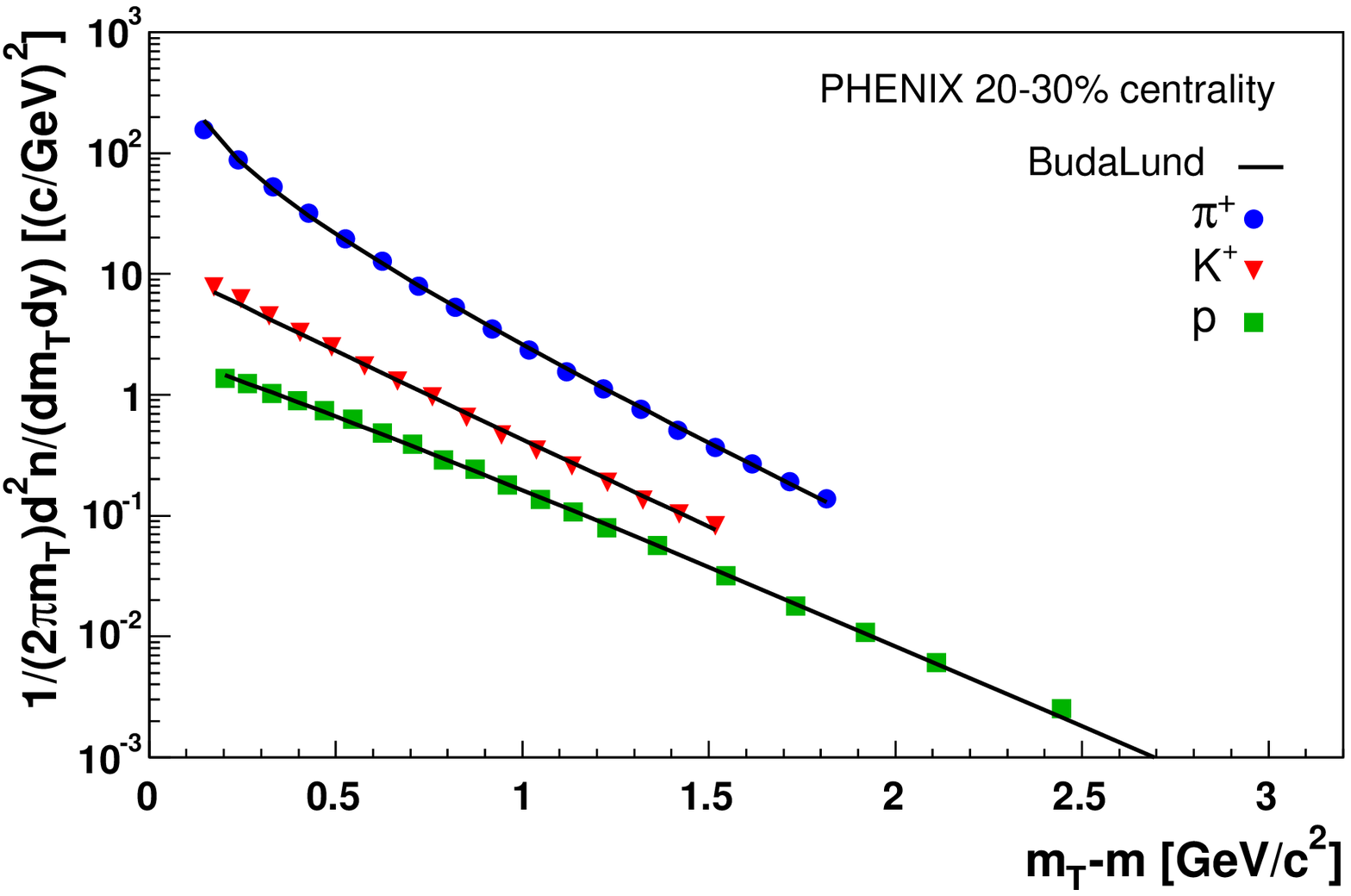}
\caption{Buda-Lund model fits to RHIC 200GeV Au+Au data
of azimuthally integrated transverse momentum spectra of 
positively charged particles~\cite{Adler:2003cb}.}
	\label{f:fig2}
\end{figure}
\begin{figure}
	\centering
		\includegraphics[width=0.45\textwidth]{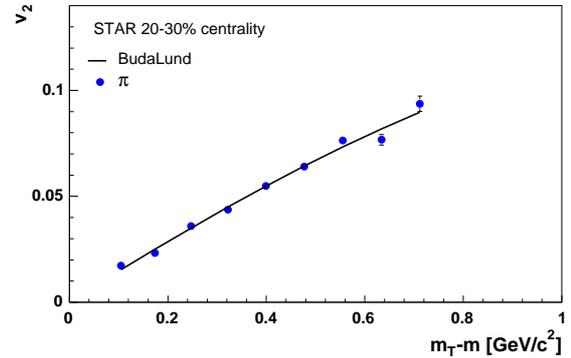}
\caption{Buda-Lund model fit to RHIC 200GeV Au+Au data
on $v_2$ elliptic flow of pions~\cite{Adams:2004bi}. }
\label{f:fig5}
\end{figure}
\begin{figure}
	\centering
		\includegraphics[width=0.45\textwidth]{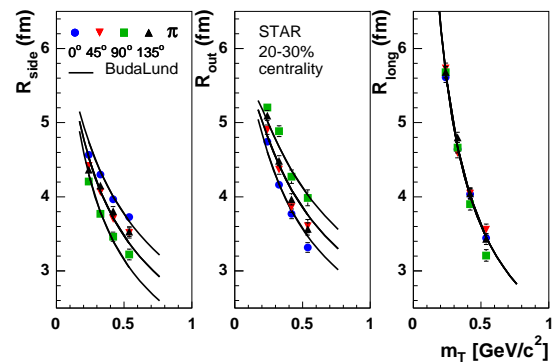}
\caption{Buda-Lund model fits to RHIC 200GeV Au+Au data~\cite{Adams:2003ra} of 
HBT radii as a function of transverse mass for different
azimuthal angles. }
	\label{f:fig3}
\end{figure}
\begin{figure}
	\centering
		\includegraphics[width=0.45\textwidth]{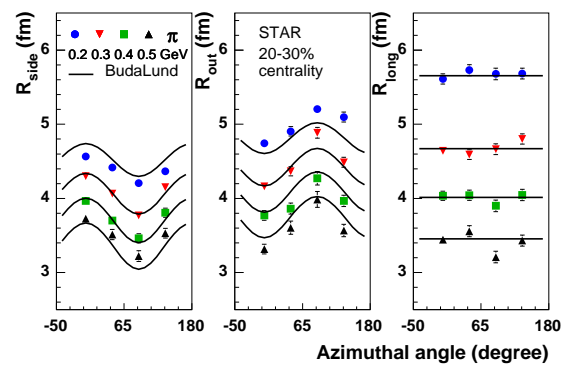}
\caption{Buda-Lund model fits to RHIC 200GeV Au+Au data~\cite{Adams:2003ra} of 
HBT radii as functions of azimuthal angles for different values of transverse mass. }
\label{f:fig4}
\end{figure}
\begin{table}
\begin{center}
\begin{tabular}{|l|l|}
\hline
\hline
Buda-Lund& \multirow{2}{*}{parameter description} \\ 
parameters&\\
\hline
\hline
$T_0$          & Temperature in the center \\
$a_\tau$       & Temperature gradient in proper time $\tau$ \\
$a_x$          & Temperature gradient in direction $x$\\
$a_y$          & Temperature gradient in direction $y$ \\\hline 
$\Delta\tau$   & Distribution width in proper-time $\tau$\\
$\Delta \eta $ & Distribution width in space-time rapidity $\eta$ \\
$R_x$          & Geometrical size in direction $x$ \\
$R_y$          & Geometrical size in direction $y$ \\\hline
$H_x$          & Expansion strength in direction $x$ \\
$H_y$          & Expansion strength in direction $y$ \\\hline
$\tau_0$       & Mean freeze-out proper-time \\\hline
$\mu_B$        & Baryochemical potential, $1/2 \  (\mu_{0,p}-\mu_{0,\overline{p}})$\\
\hline
\hline
\end{tabular}
\end{center}
\caption
{
Description of the parameters of the model. The values are meant at the mean freeze-out propoer-time $\tau_0$.
}
\label{t:pars}
\end{table}

\begin{table}[b]
\begin{center}
\begin{tabular}{|l|rl|rl|}
\hline
\hline
Buda-Lund     &\multicolumn{2}{c|}{Au+Au \@ 200 GeV}&\multicolumn{2}{c|}{Au+Au \@200 GeV} \\ 
parameters    &\multicolumn{2}{c|}{central (0-30\%)}&\multicolumn{2}{c|}{non-central (20-30\%)} \\
\hline
\hline
$T_0$ [MeV]          & 196      &$\pm$ 13   & 174    &$\pm$ 6 \\
$T_e$ [MeV]          & 117      &$\pm$ 11   & 130    &$\pm$ 6 \\ 
$\mu_B$ [MeV]        &  31      &$\pm$ 28   &  27    &$\pm$ 16 \\\hline
$R_x$ [fm]           & 13.5     &$\pm$ 1.7  & 9.5    &$\pm$ 0.5 \\
$R_y$ [fm]           & $R_x$    &           & 7.0    &$\pm$ 0.2 \\
$R_{sx}$ [fm]        & 12.4     &$\pm$ 1.6  & 12.8   &$\pm$ 0.8 \\
$R_{sy}$ [fm]        & $R_{sx}$ &           & 16.9   &$\pm$ 1.6 \\
$H_x$                & 0.119    &$\pm$ 0.020& 0.158  &$\pm$ 0.002 \\
$H_y$                & $H_x$    &           & 0.118  &$\pm$ 0.002 \\\hline
$\tau_0$ [fm/c]      & 5.8      &$\pm$ 0.3  & 5.4    &$\pm$ 0.1 \\
$\Delta\tau$ [fm/c]  & 0.9      &$\pm$ 1.2  & 2.5    &$\pm$ 0.2 \\
$\Delta\eta$         & 3.1      &$\pm$ 0.1  & 2.5    &$\pm$ 0.3 \\
\hline
$\chi^2/$NDF  &\multicolumn{2}{c|}{114/208 \@ = 0.55} &\multicolumn{2}{c|}{269.4/152 \@ = $1.77$} \\
\hline
\hline
\end{tabular}
\end{center}
\caption
{
Source parameters from simultaneous fits to PHENIX and STAR data of
Au+Au collisions at $\sqrt{s_{NN}} = 200 $ GeV,
as given in Fig. 1-5, obtained by the Buda-Lund model.
For non-central data the value of $\chi^2$/NDF refers to fits 
with statistical errors, only.
}
\label{t:results}
\end{table}

The general observation is that the Buda-Lund model parameters 
describing the source of non-central reactions
are usually slightly smaller then those of more central
collisions. However, the changes are usually within 2 standard deviations,
therefore the above statement is based on the tendency of the parameters,
and on some lower energy results not shown here but presented in 
ref.~\cite{Csanad:2004mm}, too.
For example, the central temperature in these particular
non-central reactions is below that of the more central ones. Also,
the transverse geometrical radii at the mean emission time are
considerably smaller compared to the more central values. 
Moreover, the geometric shape evolution due to 
the asymmetric particle transverse flow in-plane (x) an out-of-plane (y)
directions results in a source more elongated in-plane. Due to the
smaller longitudinal source size, the 
parameter corresponding to the formation of hydrodynamic phase is 
about 10\% \ smaller than that in more central collisions, 
$\tau_0 (20-30\%) = 5.4 \pm 0.1$~fm/$c$. The elongation in longitudinal
direction is similarly smaller, $\Delta\eta (20-30\%)= 2.5 \pm 0.3$. 
In both cases the baryochemical potential is found to be at a low
level with respect to the proton mass. We emphasize again that
the observations are based on all the fit results in 
ref.~\cite{Csanad:2004mm}. 

Note that some of the azimuthally sensitive data have large
systematic errors that affect the success of fits which we had to take 
into account. The reason for that is the difficulty of precise 
determination of the event reaction plane the data are relative to. 
Several methods are used by the experiments to overcome it and we 
mention those applied for the selected data. 

The data set we used for fitting $v_2$ was calculated by the 
four-particle cumulants reaction plane determination method that is based on 
calculations of $N$-particle correlations and non-flow effects
subtracted to first order when $N$ is greater than 2. The higher $N$ is the
more precise the event plane determination is, as expected.
STAR published two-particle cumulants $v_2$ data in the same reference, too,
but because of the visible deviations between the two kinds of data sets and 
with respect to the comments above we used $v_2\{4\}$ data, only. 
For further details, see ref.~\cite{Adams:2004bi}.

In case of azimuthally sensitive correlation radii, STAR has cast 
about 10\% possible systematic errors on the data on average. The most
likely deviations were assumed to take effect on the 'side' and 'out' radii
of transverse momentum of 0.2 GeV/$c$. The $\chi^2$/NDF for the full fit,
including HBT radii with their statistical errors is 269.6/152, 
which corresponds to a very low confidence level. 
But, when we tested our fits with the above mentioned two radii of 'side' and 'out' 
of transverse momentum of 0.2 GeV/$c$ shifting them within their 
systematic errors (about $\pm$ 5\%) we could achieve an
acceptable 1\% confidence level for the full simultaneous fit.
However, without the contribution of the HBT radii to $\chi^2$/NDF 
the confidence level is of an acceptable level of 5.1\%. 

For comparison to results in refs.~\cite{Retiere:2003kf,Tomasik:2004bn} of the 
azimuthally sensitive extension of the Blast Wave model, 
we have calculated the derivative values given in eqs.~(\ref{e:rho0}, \ref{e:rho2})
as $\rho_0 = 1.16 \pm 0.05$ and $\rho_2 = 0.29 \pm 0.02$.



\section{Conclusions}
\label{conc}

We have presented the extension of the Buda-Lund hydrodynamic model
from central ultra-relativistic heavy ion collisions to peripheral ones.
Spectra and the elliptic flow of identified particles have been described along
with the azimuthal dependence of two-particle correlation function
radii in the ellipsoidally symmetric generalization of the model.
Theoretical predictions were tested against RHIC data. From model fits
to data of 20-30\% centrality class at mid-rapidity source parameters characterizing
these non-central ultra-relativistic heavy ion reactions were extracted.\\

The results of our analysis indicate that the central temperature in 20-30\% 
centrality reactions is lower than that in more central collisions, 
$T_{0} = 174 \pm 6$~(stat)~MeV. In an earlier analysis we showed that in more 
central (0-30\%) reactions of the same collision energy this value was 
$T_{0} = 196 \pm 13$~(stat)~MeV. 
We have found that the transverse flow is stronger in the 
reaction plane than out of plane
with Hubble constants $H_x=0.158 \pm 0.002$ and $H_y=0.118 \pm 0.002$.
The almond shape of the reaction zone initially elongated 
out of plane gets slightly elongated in the direction of the impact parameter 
by the time the particle emission rate reaches its maximum. The effect is 
reflected by the geometrical radii in the two perpendicular directions at 
that time, $R_x(\mathrm{in-plane}) = 9.5 \pm 0.5$~(stat)~fm,  $R_y(\mathrm{out-plane})
= 7.0 \pm 0.2$~(stat)~fm. This is the first time that an in-plane extended
source has been reconstructed from simultaneous hydrodynamic model fits 
to identified particle spectra, elliptic flow and azimuthally sensitive 
HBT data in 200 AGeV Au+Au collisions at RHIC.

The qualitative agreement between model and data is apparently
good as can be judged from the figures. Although, in these fits data 
were used with statistical errors, only, which resulted in $\chi^2/$NDF$ = 1.77$.


\paragraph*{Acknowledgments}
Inspiring discussions with professors R. J. Glauber at Harvard University
and Mike Lisa at the WPCF 2009 conference at CERN are gratefully acknowledged.
This paper was prepared within a bilateral collaboration
between Hungary and Slovakia under project Nos.\ SK-20/2006 (HU), SK-MAD-02906 (SK).
B.T.\ acknowledges support by MSM 6840770039 and
LC 07048 (Czech Republic). M.Cs.,\ T.Cs.\ and A.S.\ gratefully acknowledge the support of
the Hungarian OTKA  grants T49466 and NK 73143. T.  Cs. was supported also by a 
Senior Leaders and Scholars Fellowship of the Hungarian American Enterprise Scholarship Fund.

\bibliographystyle{prlsty}

\end{document}